\documentclass[aps,twocolumn,showpacs]{revtex4}

\usepackage{color}
\usepackage{graphicx}
\usepackage{epstopdf}
\usepackage{graphics}
\usepackage{amssymb}
\usepackage{amsmath}
\usepackage{latexsym}
\usepackage{color}
\usepackage{subfigure}

\newcommand{\tr}{{\rm tr \thinspace}}
\newcommand{\bra}[1]{\left\langle{#1}\right\vert}
\newcommand{\ket}[1]{\left\vert{#1}\right\rangle}
\def\ketc[#1]{\vert #1 \rangle}
\def\brac[#1]{\langle #1 \vert}

\newcommand{\expect}[1]{\langle{#1}\rangle}

\newcommand{\abs}[1]{\left\vert{#1}\right\vert}

\newcommand{\cnote}{\mbox{CNOT}_{e}}
\newcommand{\cnotn}{\mbox{CNOT}_{n}}

\newcommand{\beq}{\begin{equation}}
\newcommand{\eeq}{\end{equation}}
\newcommand{\bqa}{\begin{eqnarray}}
\newcommand{\eqa}{\end{eqnarray}}
\newcommand{\nn}{\nonumber}

\newcommand{\erf}[1]{Eq.~(\ref{#1})}
\newcommand{\dg}{^\dagger}
\newcommand{\bsig}{{\mbox{\boldmath $\sigma$}}}

\begin{document}

\title{Quantum non-demolition measurements of single donor spins in semiconductors}

\author{Mohan Sarovar$^{1}$}
\author{Kevin C. Young$^{1,2}$}
\author{Thomas Schenkel$^{3}$}
\author{K. Birgitta Whaley$^{1}$}
\affiliation{Berkeley Center for Quantum Information and Computation, Departments of Chemistry$^{1}$ and Physics$^{2}$, University of California, Berkeley, California 94720 \\
$^{3}$Accelerator and Fusion Research Division, Lawrence Berkeley National Laboratory, Berkeley, California 94720}


\begin{abstract}
We propose a technique for measuring the state of a single donor electron spin using a field-effect transistor induced two-dimensional electron gas and electrically detected magnetic resonance techniques. The scheme is facilitated by hyperfine coupling to the donor nucleus. We analyze the potential sensitivity and outline experimental requirements. Our measurement provides a single-shot, projective, and quantum non-demolition measurement of an electron-encoded qubit state. 
\end{abstract}
\pacs{73.23.-b, 03.67.Lx, 76.30.-v, 84.37.+q}


\maketitle


Semiconductor implementations of quantum computation have become a vibrant subject of study in the past decade because of the promise quantum computers (QCs) hold for radically altering our understanding of efficient computation, and the appeal of bootstrapping the wealth of engineering experience that the semiconductor industry has accumulated. A promising avenue for implementing quantum computing in silicon was proposed by Kane \cite{Kan-1998}, suggesting the use of phosphorous nuclei to encode quantum information. However, while the long coherence times of the nuclei are advantageous for information storage tasks, their weak magnetic moment also results in long gate operation times. In contrast, donor \emph{electrons} in Si couple strongly to microwave radiation and permit the fast execution of gates; and while electron spin decoherence times are shorter than their nuclear counterparts, the tradeoff of decreased robustness to noise for faster operation times could be appropriate to implementing a fault-tolerant QC. This has led several authors to suggest the use of electron spin qubits as a variant on the original Kane proposal (e.g. \cite{Vri.Yab.etal-2000, Hil.Hol.etal-2005, Sch.Lid.etal-2006}), and we focus on such a modified Kane architecture here. 

An integral part of any quantum computation architecture is the capacity for high-fidelity qubit readout.  While small ensembles of donor spins have been detected \cite{McC.Hue.etal-2006, Ste.Boe.etal-2006} and single spin measurements have been demonstrated (e.g. \cite{Elz.Han.etal-2004, Xia.Mar.etal-2004}), detection of spin states of single donor electrons and nuclei in silicon has remained elusive.  In this paper we analyze spin dependent scattering between conduction electrons and neutral donors \cite{Gho.Sil-1992, Lo.Bok.etal-2007} as a spin-to-charge-transport conversion technique, and show that quantum non-demolition (QND) measurements of single electron spin-encoded qubit states are realistically achievable when mediated via nuclear spin states.  Such a measurement will also be of value to the developing field of spintronics \cite{Zut.Fab.etal-2004} where the electrical detection of spin states is valuable. Our readout takes advantage of two features: i) the ability to perform electron spin resonance spectroscopy using a two-dimensional electron gas (2DEG), and ii) the hyperfine shift induced on dopant electron Zeeman energies by the dopant nuclear spin state. 

In the next section we describe the experimental apparatus and the techniques of 2DEG mediated spin resonance spectroscopy. In section \ref{sec:proposal} we present our proposal for spin state measurement in detail, and then in section \ref{sec:sens} we analyze the sensitivity of the measurement scheme and establish the key factors that determine signal-to-noise. Then section \ref{sec:conc} concludes with a discussion.

\section{The physical setting}
\label{sec:exp}

\begin{figure}[h!]
\includegraphics[scale=0.095]{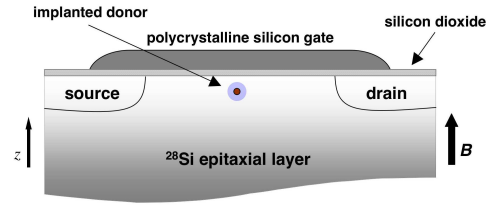}
\caption{A cross section of the field-effect transistor (FET) used to create the 2DEG. In order to reduce qubit decoherence, it is beneficial to implant into isotopically purified silicon.} \label{fig:cross_section}
\end{figure}

The use of electrical conductivity properties of semiconductors to investigate spin properties of (bulk-doped) impurities has a long history \cite{Hon-1966, Sch.Sol-1966}, including studies of donor polarization using a 2DEG probe \cite{Gho.Sil-1992}. Figure \ref{fig:cross_section} shows a cross section of a 2DEG spin readout device with a single implanted donor. Prior studies have used similar devices with bulk-doping \cite{Gho.Sil-1992, Wil.Hue.etal-2008} or a large number of implanted donors ($~10^6$) \cite{Lo.Bok.etal-2007} in the 2DEG channel. The 2DEG is operated in accumulation mode and thus conduction electrons scatter off the electron(s) bound to the shallow donor(s). The basic principle exploited in these studies is the role of the exchange interaction in electron-electron scattering.  At a scattering event between a conduction electron and a loosely bound donor impurity electron, the Pauli principle demands that the combined wave function of the two electrons be antisymmetric with respect to coordinate exchange. This constraint, together with the fact that the combined spin state can be symmetric (triplet) or antisymmetric (singlet), imposes a correlation between the spatial and spin parts of the wave function and results in an effective spin dependence of the scattering matrix, leading to a spin dependent conductance. Application of a static magnetic field will partially polarize conduction and impurity electrons leading to excess triplet scattering. A microwave drive will alter these equilibrium polarizations when on resonance with impurity (or conduction) electron Zeeman energies and hence alter the ratio of singlet versus triplet scattering events, registering as a change in the 2DEG current. Thus, the spin dependent 2DEG current can be used as a detector of spin resonance and accordingly this technique is commonly known as \textit{electrically detected magnetic resonance} (EDMR). Ghosh and Silsbee, and later Willems van Beveren \textit{et. al.}, employed EDMR in bulk doped natural silicon to resolve resonance peaks corresponding to donor electron spins that are hyperfine split by donor P nuclei \cite{Gho.Sil-1992, Wil.Hue.etal-2008}. Recently, Lo \textit{et. al.} have used this technique  to investigate spin dependent transport with micron-scale transistors on isotopically enriched $^{28}$Si implanted with $^{121}$Sb donors \cite{Lo.Bok.etal-2007}.

\section{The proposal: EDMR based single spin measurement}
\label{sec:proposal}

From hereon we will consider the experimental setup described above in the particular situation where there is a single donor P nucleus present, the electron spin of which encodes the quantum information that we wish to measure. A crucial question in the context of quantum computing is whether the spin-dependent 2DEG current can be used to measure the state of an electron-spin qubit, as spin-dependent tunneling processes have been employed \cite{Elz.Han.etal-2004, Xia.Mar.etal-2004}. The fundamental concern here is whether the spin exchange scattering interaction at the core of the spin-dependent 2DEG current allows for a quantum state measurement of a single donor impurity electron spin. 

A spin ($1/2$) state measurement couples the microscopic state of the spin, given in general by a (normalized) density matrix, 
\beq
\rho_i = \left(\begin{array}{cc}a & c \\c^* & b\end{array}\right)
\eeq
(in the measurement basis, with $a+b=1$), to a macroscopic meter variable $I$, the 2DEG current in our case. The meter variable can take one of two values, and at the conclusion of the measurement, a faithful measuring device would register each meter variable with the correct statistics, i.e., $I_{\uparrow}$ with probability $a$ and $I_{\downarrow}$ with probability $b$. A QND measurement device will have the additional property that once a meter variable has been registered, the measured spin remains in the state corresponding to the value registered so that a second measurement gives the same result \cite{Bra.Vor.etal-1980, Bra.Kha-1995}.

One might expect that because the exchange interaction is destructive (in the sense that it will change the state of the target (donor electron) spin with some probability), it will only produce a faithful measurement if the time over which it acts is extremely short. We will now show that this is indeed the case and that direct measurement of the electron spin via the 2DEG current is consequently not possible within experimentally realizable times.  This negative result will motivate our subsequent presentation in Section~\ref{sec:proposal_2} of a more complex scheme for measurement that is both faithful and experimentally realizable. 

\subsection{Direct measurement of electron spin}
\label{sec:proposal_1}

To investigate the ability of the spin-dependent 2DEG current to measure the electronic spin state, 
we shall use a minimal model of the scattering process. Since we are primarily concerned with the spin state of the particles involved in the scattering, we examine the transformation that a single scattering event induces on the spinor components of the conduction and impurity electrons. We write this transformation as 
\beq
\rho_{\textrm{out}}(\bf{k},\bf{k'}) = \frac{\mathcal{T}_{\bf{k},\bf{k}'} \rho_{\textrm{in}} \mathcal{T}_{\bf{k},\bf{k}'}\dg}{\tr(\mathcal{T}_{\bf{k},\bf{k}'} \rho_{\textrm{in}} \mathcal{T}_{\bf{k},\bf{k}'}\dg)}, 
\eeq
where $\rho_{\textrm{in/out}}$ are the density operators for the spin state of the combined two-electron system, and: $\mathcal{T}_{\bf{k},\bf{k}'} = F_\textrm{d}({\bf k},{\bf k}') + F_\textrm{x}({\bf k},{\bf k}') \bsig_c \cdot \bsig_i$ \cite{Gla-1963}. Here $F_\textrm{d}$ ($F_\textrm{x}$) is the amplitude for un-exchanged (exchanged) conduction and impurity electron scattering \footnote{We assume throughout that the operating temperature is above the Kondo temperature of the device.}. Note that the spatial aspects of the problem only enter into the amplitudes. We will assume elastic scattering with the donor electron remaining bound, and no scattering of conduction electrons outside the 2DEG. Therefore the amplitudes can be parameterized by two parameters: $F_{\textrm{d/x}} \equiv F_{\textrm{d/x}}(\theta, k)$, the scattering angle within the 2DEG,  $\theta$; and the incoming momentum magnitude $k$ (determined by the Fermi energy of the 2DEG electrons). These amplitudes are free parameters in our model and we explore a wide range of values for them in the simulations below. The direct and exchange scattering amplitudes are simply related to the more familiar singlet ($f_s$) and triplet ($f_t$) scattering amplitudes as:
\bqa
F_d(\theta, k) &=& \frac{1}{4}\left( f_s + 3f_t \right) \nn \\
F_x(\theta, k) &=& \frac{1}{4}\left( f_t - f_s \right)
\eqa


Now, assume an initial state $\rho_{\textrm{in}} = (p\ket{\uparrow}_c\bra{\uparrow} + (1-p)\ket{\downarrow}_c\bra{\downarrow}) \otimes \rho_\textit{i}$, where the first term in the tensor product is the state of the conduction electron (the conduction band is assumed to be polarized to the degree $P_c^0=2p-1, ~0\leq p \leq 1$), and the second term is the general state of the donor electron given above. After applying the scattering transformation and tracing out the conduction electron (because we have no access to its spin after the scattering event in this experimental scheme) we get a map that represents the transformation of the impurity electron state due to one scattering event:
\bqa
&\rho_i & \rightarrow \rho_i '(\theta, k) \nn \\
&=& \frac{(1-p)}{\mathcal{N}}\left[ (F_\textrm{d} + F_\textrm{x}\mathbf{\sigma}_z)\rho_i (F^*_\textrm{d} + F^*_\textrm{x}\mathbf{\sigma}_z) + 4|F_x|^2 \sigma_- \rho_i \sigma_+ \right] \nn \\ 
&& + \frac{p}{\mathcal{N}}\left[ (F_\textrm{d} - F_\textrm{x}\mathbf{\sigma}_z)\rho_i (F^*_\textrm{d} - F^*_\textrm{x}\mathbf{\sigma}_z) + 4|F_x|^2 \sigma_+ \rho_i \sigma_- \right] \nn \\
\label{eq:trans}
\eqa
where $\mathcal{N}$ is a normalization constant to ensure $\tr(\rho_i ')=1$, and we have not explicitly written the $(\theta, k)$ dependence of the scattering amplitudes for brevity. 

In order for the measurement to be faithful, the diagonal elements of the impurity spin state (the population probabilities) must be preserved under the interaction -- that is, the measurement interaction may induce dephasing (in the measurement basis), but no other decoherence. However, the terms proportional to $|F_\textrm{x}|^2$ in \erf{eq:trans} suggest that there will be population mixing. Using this equation, we can write the transformation of the diagonal elements (which are uncoupled from the off diagonal elements by the transformation \erf{eq:trans}) as:
\bqa
&a& \rightarrow a'(\theta, k) \nn \\
&=& \frac{(1-2P_c^0\Lambda \cos\chi + (2P_c^0-1)\Lambda^2)a + 2(1-P_c^0)\Lambda^2 }{4P_c^0\Lambda(\Lambda - \cos\chi)a + 1 + 3\Lambda^2 + 2P_c^0\Lambda\cos\chi - 2P_c^0\Lambda^2} \nn
\label{eq:transa}
\eqa
where $\Lambda(\theta, k) \equiv |F_x(\theta, k)|/|F_d(\theta,k)|$, and $\chi(\theta, k) \equiv \arg F_x(\theta,k) - \arg F_d(\theta, k)$, and we have used the fact that $a+b=1$ to normalize the transformation.

We can iterate this recursion to simulate the effects of the repeated scattering events that contribute to the current. An appropriate quantification of measurement quality is the \textit{measurement fidelity} \cite{Ral.Bar.etal-2006}: 
\beq
\mathcal{F}_n = 2|(\sqrt{a^{(n)}}\sqrt{a^{(0)}} + \sqrt{b^{(n)}}\sqrt{b^{(0)}})^2-0.5|, 
\eeq
where $a^{(n)}$ and $b^{(n)}$ are the diagonal elements of $\rho_i$ after $n$ scattering events. An ideal measurement has $\mathcal{F}_n=1$, while $\mathcal{F}_n=0$ indicates a measurement that yields no information  -- i.e. no correlation between the original qubit state and the meter variables. Since the measurement should work for all initial states, we consider the worst-case measurement fidelity: $\mathcal{F}^w_n = \min_{a^{(0)},b^{(0)}}\mathcal{F}_n$. 

\begin{widetext}

\begin{figure}[t!]
\includegraphics[scale=0.54]{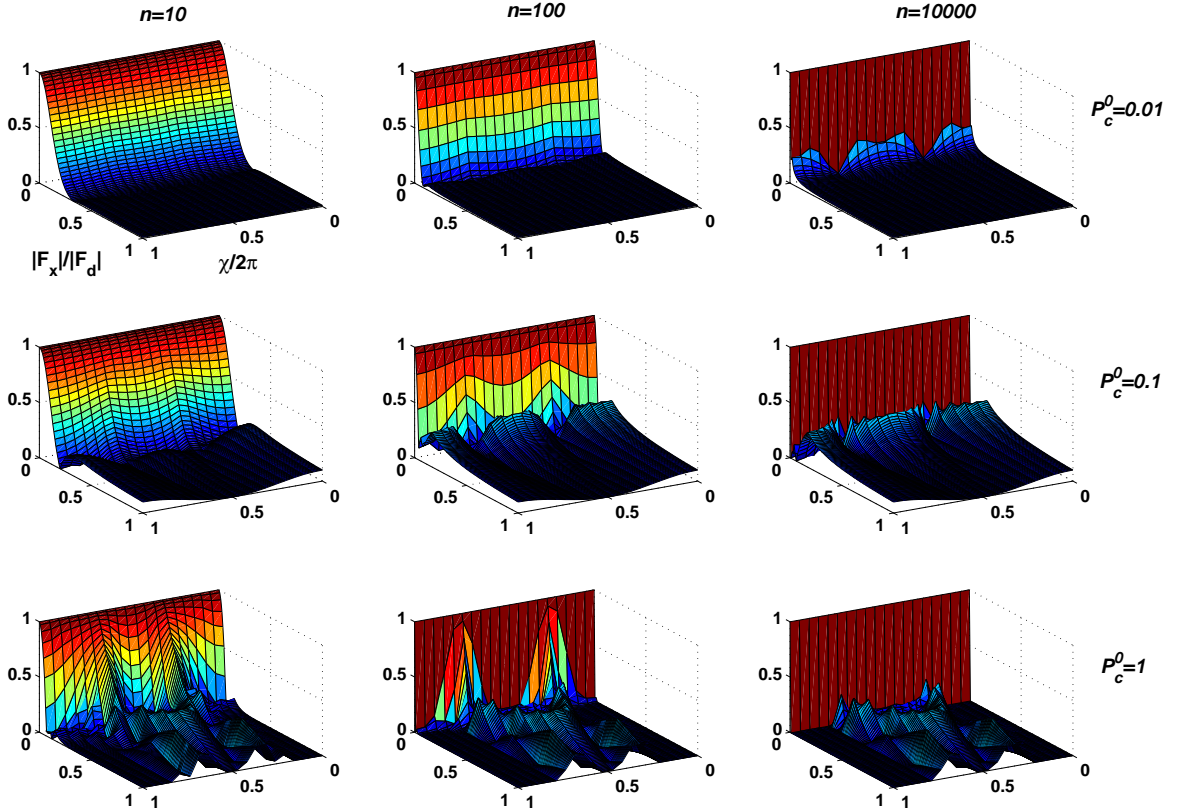}
\caption{Evolution of worst-case measurement fidelity, $\mathcal{F}^w_n$, during 2DEG scattering dynamics as a function of the number of scattering events $n$, for a range of values of the ratio of direct and exchange scattering amplitudes and of 2DEG equilibrium polarization. The independent (base plane) axes on the plots parametrize the complex scattering amplitude ratio $F_x/F_d \equiv |F_x|/|F_d|e^{i\chi}$: one axis is the magnitude, $\Lambda \equiv |F_x|/|F_d|$ (shown for $0 < |F_x|/|F_d| < 1$; the plots are restricted to this range because the worst-case fidelity is negligibly small outside it), and the other is the phase, $\chi$ (shown for $0 <  \chi/2\pi < 1$). The number of scattering events, $n$, varies across the columns, with values $n=10, 10^2$ and $10^4$ shown here. The 2DEG equilibrium polarization, $P_c^0$, varies across the rows, with values $P_c^0=0.01,0.1$ and $1$ shown here. For $P_c^0\geq 0.1$, we see that there are fairly large regions in the $F_x/F_d$ parameter space for which the worst-case measurement fidelity is non-zero: however, (i) $\mathcal{F}^w_n$ still decays rapidly with number of scattering events, and is rarely $>0.9$ (the fidelities desirable for high-quality measurement), and (ii) $\mathcal{F}^w_n$ is highly sensitive to the precise value of $F_x/F_d$ and $P_c^0$ in these regions.} \label{fig:worst_case_fids}
\end{figure}

\end{widetext}

In order to calculate this fidelity we need to determine how many scattering events will take place within the time required to do the measurement. Given the current state of the art, and factoring in improvements in 2DEG mobility \cite{Eng.McF.etal-2007} and conduction electron polarization, we estimate a shot-noise limited measurement time of $\tau_m \sim 10^{-3} ~\textrm{s}$ (this calculation is given below, in section \ref{sec:sens}). Within this time, there will be $\sim 10^{9}$ scattering events (see the Appendix for details on calculating the number of scattering events per second). Although we do not assume specific values of the scattering amplitudes, we find from iterating the above recursion for a broad range of values $F_x/F_d$ that after $\sim 10^{9}$ scattering events $\mathcal{F}^w_n$ is $\ll 1$ for any non-zero value of the exchange amplitude $|F_x|$ and for any polarization, $P_c^0$. Figure \ref{fig:worst_case_fids} shows worst-case fidelity decay as a function of scattering amplitude parameters for various values of 2DEG polarization $P_c^0$. These simulations clearly show that the relaxation of a general electron spin state is rapid across virtually all reasonable parameter ranges. In fact, for realistic 2DEG polarization values $\mathcal{F}^w_n$ typically drops to near zero already after $\sim 10^3-10^4$ scattering events. Thus the measurement induced population mixing time is $T_{\textrm{mix}} \sim 1-10 \textrm{ns}$, which is drastically smaller than $\tau_m$. These simulations thus show conclusively that whatever the precise values of the scattering amplitudes, under realistic experimental conditions the electron spin relaxation induced by the scattering interaction makes the 2DEG current an ineffective measurement of the electron spin state. This makes it impossible to faithfully map the electron spin state onto the meter variable, and hence impossible to perform a single electron spin state measurement using the 2DEG current directly.

For completeness we note that since we only have access to the total 2DEG current and no angle-resolving detectors, the actual impurity electron density matrix must also involve an average over $\theta$ and $k$ over the 2DEG Fermi surface in \erf{eq:trans}. However, as we have shown that the direct measurement will not work for any value of $\theta$ and $k$, the averaged dynamics will only result in a worse performance analysis.

However, as we will now show, it is possible to make use of the nuclear spin degree of freedom in order to utilize EDMR for projective and QND measurement of single spin states. The key is that the state of the nuclear spin affects the Zeeman splitting of the electron spin (and thus its resonant frequency) via the mutual hyperfine coupling. Therefore our strategy is to transfer the qubit state from the electron to the nucleus and then to perform an EDMR readout.

\subsection{Nuclear spin mediated electron spin state measurement}
\label{sec:proposal_2}

\begin{figure}
\begin{center}
\includegraphics[scale=0.35]{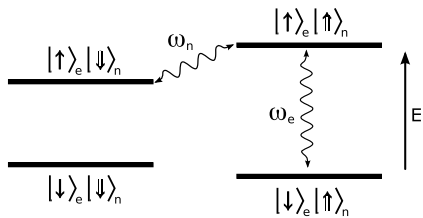}
\end{center}
\caption{Four-level system of electron-nuclear spin degrees of freedom.  The energy eigenstates in the secular approximation are the eigenstates of $\sigma_z^e$  and $\sigma_z^n$. The transitions indicated by arrows are required for the state transfer described in the text. \label{fig:estates}}
\end{figure}

The low-energy, low-temperature Hamiltonian describing the electron and nuclear spins of a phosphorous dopant in a static magnetic field, $\mathbf{B} = B \hat{\mathbf{z}}$ is 
\beq
H = \frac{1}{2} \left[ g_e \mu_B B \sigma_z^e - g_n \mu_n B \sigma_z^n \right] + A \bsig^e \cdot \bsig^n 
\label{eq:ham}
\eeq
where $\mu_B$ and $\mu_n$ are the Bohr and nuclear magnetons, $g_e$ ($g_n$) is the electron (nuclear) \textit{g}-factor, and $A$ characterizes the strength of the hyperfine interaction between the two spins \cite{Kan-1998} (we set $\hbar=1$ throughout the paper). For moderate and large values of $B$, the $\sigma_z$ terms dominate and we can make the \textit{secular approximation}, to arrive at: $H \approx 1/2 \left[ g_e \mu_B B \sigma_z^e - g_n \mu_n B \sigma_z^n \right] + A \sigma_z^e \sigma_z^n$. The energy levels and eigenstates of this Hamiltonian are shown in Fig. \ref{fig:estates}. Note that we have ignored the coupling of both spins to uncontrolled degrees of freedom such as paramagnetic defects and phonons (coupling to lattice spins can be mitigated by the use of a $^{28}$Si substrate). These environmental  couplings will contribute to decoherence of the nuclear and electron spin states (e.g. \cite{Sou.Das-2003}), and we will simply assume that this results in some effective relaxation and dephasing of the electron and nuclear spins.

We see that the resonance frequency (Zeeman energy) of the electron is a function of the nuclear spin state. Therefore, our strategy will be to transfer the qubit state from the electron to the nucleus and then use EDMR to measure the nuclear spin. This is in effect a spin-to-resonance-to-charge conversion measurement.

To perform the state transfer, we appeal to the qubit SWAP gate: $\mbox{SWAP} [ \rho_e \otimes \tau_n] \mbox{SWAP}\dg = \tau_e \otimes \rho_n$. SWAP can be decomposed into the sequence of three controlled-not (CNOT) gates \cite{mikeandike} $\mbox{SWAP} = \cnotn \cnote \cnotn $, where the subscript indicates which of the two qubits is acting as the control.  However, the complete exchange of electron-nuclear states is unnecessary, since the spin state of the impurity electron is lost to the environment by the application of resonant pulses and elastic scattering with conduction electrons in the 2DEG.  Therefore, the final operator in the sequence can be neglected since it only alters the state of  the electron. This leads  to the definition of the electron-to-nucleus transfer gate, $\mbox{TRANS}_e = \mbox{CNOT}_e \mbox{CNOT}_n$. To apply these CNOT gates we use resonant pulses: $\cnote$ interchanges the states $\ket{\uparrow}_e\ket{\Uparrow}_n$ and $\ket{\uparrow}_e\ket{\Downarrow}_n$ and so can be implemented by application of a resonant $\pi$-pulse at frequency $\omega_{n}$ (see Fig. \ref{fig:estates}), an RF transition; similarly, $\cnotn$ interchanges $\ket{\uparrow}_e\ket{\Uparrow}_n$ and $\ket{\downarrow}_e\ket{\Uparrow}_n$ and is implemented by a resonant $\pi$-pulse at $\omega_{e}$, a microwave transition. Each of these transitions is dipole-allowed, ensuring that gate times are sufficiently fast. The ability to apply pulses faster than relevant decoherence times is required for successful implementation of the state transfer. We note that after this work was completed, a state swap scheme very similar to the state transfer scheme outlined above was successfully performed in experiments on bulk-doped Si:P samples \cite{Mor.Tyr.etal-2008}. 

Suppose the electron is in an initial (pure) state, $\ket{\psi}_e = \alpha\ket{\uparrow}_e + \beta\ket{\downarrow}_e$, while the nucleus is in a general mixed state, 
\beq
\tau_n = \left( \begin{array}{cc}
u & w \\
w^* & v \end{array} \right).
\eeq
After performing the state transfer on the combined state and tracing over the electron degrees of freedom (because it is lost to the environment), we are left with the reduced density matrix describing the nucleus, 
\bqa
\tr_e\left(\mbox{TRANS}_e \left[ \rho_e \otimes \tau_n \right]\mbox{TRANS}_e^\dagger \right) = \\ \nn 
\left( \begin{array}{cc} |\alpha|^2 &  \alpha \beta^*(w+w^*) \\
 \alpha^* \beta (w+w^*) & |\beta|^2  \end{array} \right)
\eqa
 Because of the hyperfine coupling (Fig. \ref{fig:estates}), electron resonance will occur at the lower frequency with probability $\abs{\alpha}^2$ and at the higher frequency with probability $\abs{\beta}^2$. The electrical detection of this shift from the free electron resonance frequency by EDMR constitutes a single-shot, projective measurement in the $\sigma_z$ basis of the original electron state (and therefore, qubit state) with the correct statistics.

In detail, the single qubit spin readout can proceed as follows. Following state transfer to the nuclear spin, one of the two hyperfine split electron spin resonance lines that corresponds to a given nuclear spin projection is addressed by dialing in the corresponding microwave frequency for resonant excitation of electron spin transitions.  At the same time, the transistor is turned on and the channel current is monitored.  Now, assume that the magnetic fields have been tuned to address the $\ket{\uparrow}_n$ nuclear state projection. Then with probability $|\alpha |^2$ the transistor current will differ from the off-resonant current value and with probability $|\beta |^2$ it will be just equal to the off-resonant channel current. In either case, monitoring the current at one hyperfine resonance for the $\tau_m$ measurement duration constitutes a readout of the nuclear spin. And due to the prior state transfer, it effectively measures the spin state of the original donor electron spin.  

\begin{widetext}

\begin{figure}[h!]
\includegraphics[scale=0.21]{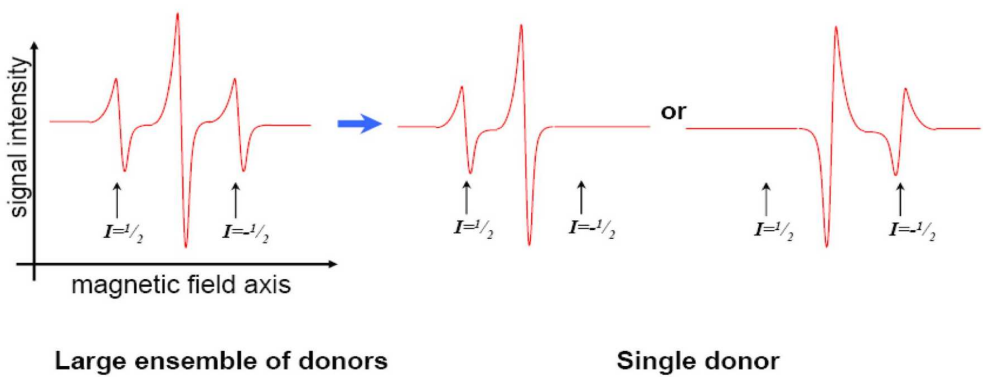}
\caption{Illustration of single spin readout.  In experiments with large ensembles of donor spin qubits, lines from all nuclear spin projections are present in EDMR measurements (left).  In measurements with single donors (right), only single lines are present for measurement times shorter then the nuclear spin relaxation time.  Monitoring the current at a given resonant field measures the spin state of the donor nucleus with the correct statistics.} \label{fig:meas_signal}
\end{figure}

\end{widetext}

\section{Measurement sensitivity and measurement-induced decoherence}
\label{sec:sens}

Two critical practical issues need to be addressed for realization of this protocol for measurement of the donor electron spin quantum state.  These are: i) the sensitivity of the EDMR measurement needs to be sufficiently high to allow single donor spins to be detected, and ii) the measurement time $\tau_m$ must be small compared to the lifetime of the nuclear spin.

We first address the issue of the sensitivity of the differential EDMR current in the limit of single donor scattering. As detailed above, the spin state of the donor electron is continually changing due to the scattering interaction, and hence is time-dependent. However, ignoring the transient, we can approximate it with a time independent value given by the steady state solution of the recursion relation, \erf{eq:trans}. This approximation can be thought of as taking the equilibrium spin value, where the ``spin temperature" of the impurity has equilibrated with that of conduction electrons via the scattering interaction. The explicit simulations of the scattering recursion \erf{eq:transa} shown in Section~\ref{sec:proposal_1} indicate that this equilibration happens within $10^4$ scattering events for all possible values of scattering amplitudes. Thus the time scale for this equilibration is $\sim 10\textrm{ns}$, much faster than the observable times scales of the measurement, justifying our use of the steady-state solution of the recursion (see the Appendix for details on calculating the number of scattering events per second). Solving for the steady-state ($\rho_i^{(n)}=\rho_i^{(n-1)} \equiv\rho_i^{ss}$), gives us a time-independent, non-resonant single donor ``polarization" equal to
\beq
\expect{\sigma_z}_i^{\textrm{ss}} \equiv \tr(\sigma_z \rho_i^{ss}) = \frac{\Lambda-\sqrt{(P_c^0)^2 \cos^2\chi + \Lambda^2(1-(P_c^0)^2)}}{P_c^0 (\Lambda - \cos\chi)}
\eeq
It should be noted that this expression for the steady state ``polarization" is not valid when $P_c^0 = 0$ or $\Lambda=0$, but neither of these limits is relevant to spin measurement. Now, we can follow the analysis of Ref. \cite{Gho.Sil-1992}, using donor ``polarization" $\expect{\sigma_z}_i^{\textrm{ss}}$, to estimate the on-resonant ($I$) and off-resonant ($I_0$) current differential (normalized) as:
\bqa
\frac{\Delta I}{I_0} \equiv \frac{I - I_0}{I_0} \approx -\alpha' ~s~ \expect{\sigma_z}_i^{\textrm{ss}} P_c^0 \frac{1/\tau_n}{1/\tau_t}.
\label{eq:singledonor_diff_curr}
\eqa
Here $\alpha'  \equiv \expect{\Sigma_s - \Sigma_t}\vert_{z=z_i}/\expect{\Sigma_s + 3\Sigma_t}\vert_{z=z_i}$, $\Sigma_s$ and $\Sigma_t$ are singlet and triplet scattering cross sections, respectively, and $\expect{\cdot}\vert_{z=z_i}$ denotes an average over the scattering region with the donor location in $z$ (see Fig. \ref{fig:cross_section}) held fixed \cite{Sou.Lo.etal-2008}. $s = 1 - (1-s_i)(1-s_c)$, and $s_i$ and $s_c$ (both between $0$ and $1$) are saturation parameters which characterize how much of the microwave power is absorbed by the impurity and conduction electrons, respectively \cite{Gho.Sil-1992}. $s_i$ is a function of the broadening at the \textit{single} donor electron resonance frequency: if we work in a regime where this broadening is minimal (as required to perform the quantum state transfer described above), $s_i \approx 1$ and thus $s \approx 1$. The final term in \erf{eq:singledonor_diff_curr} represents the ratio between impurity scattering ($1/\tau_n$) and total scattering ($1/\tau_t$) rates. We assume $1/\tau_t = 1/\tau_0 + 1/\tau_n$, where $1/\tau_0$ is the scattering rate due to all other processes (such as surface roughness scattering and Coulomb scattering by charged defects).

To estimate the expected magnitude of this current differential, we begin by considering present state of the art 2DEG mediated EDMR experiments where this current differential is $\sim10^{-7}$ (with T $\sim5 K, B\sim0.3T$, a 2DEG channel area of $160\times 20 \mu \textrm{m}^2$, probe current $1 \mu$A, and a donor density of $2\times 10^{11} \textrm{donors/cm}^2$) \cite{Lo.Bok.etal-2007}. We assume that $\alpha'$ will be similar for the single donor device as in current experiments. Then in scaling down to a single donor, the first aspect to consider is the scattering rate ratio: $\varrho \equiv \frac{1/\tau_n}{1/\tau_t}$. To first order this ratio can be kept constant if we scale the 2DEG area concomitantly with the donor number. From the channel area and density of current experiments, we extrapolate that a 2DEG area of $\sim 30\times 30 \textrm{nm}^2 $ -- well within the realm of current technology \cite{Par.Lid.etal-2004} -- would keep $\varrho$ unchanged. Optimization of donor depth might relax this size requirement \cite{Sou.Lo.etal-2008}. A higher order analysis would require detailed investigation of the device specific interface and intrinsic contributions to the other scattering processes, and hence to the channel mobility and $\tau_0$. Related to this concern, the mobility of the 2DEG channel can be improved -- e.g., by using hydrogen passivation to mitigate surface roughness at the oxide interface \cite{Eng.McF.etal-2007} -- to increase $\varrho$. We conservatively estimate a factor of 10 increase in $\Delta I/I_0$ from such improvements. The saturation parameter $s\sim 1$ for large enough microwave powers in the recent measurements \cite{Lo.Bok.etal-2007} and so does not present an area for improvement. Finally, an avenue for significant improvement in signal is to increase the conduction electron polarization, $P_c^0$, which is currently $\sim 0.1 - 1\%$. This polarization is roughly proportional to the applied static magnetic field, and therefore a factor of 10 improvement is possible by operating at $B=3T$. Additionally, spin injection techniques can be employed to achieve $P_c^0 > 10\%$ (e.g. \cite{App.Hua.etal-2007, Jon.Kio.etal-2007}), resulting in a 100-fold improvement in $\Delta I/ I_0$. Hence, by improvements in device scaling and channel mobility, and by incorporating spin injection, we estimate a realistic, improved current differential of $\Delta I/ I_0 \sim 10^{-4}$. Given this $\Delta I/ I_0$ and a probe current of $I_0 \sim 1\mu A$, to achieve an signal-to-noise (SNR) of 10 through shot-noise limited detection we require a $\tau_m$ satisfying: 
\beq
\left(\frac{\Delta I}{I_0}\right) \frac{I_0\tau_m}{e} > 10 \sqrt{\frac{I_0\tau_m}{e}}
\eeq
where the left hand side is the signal, the right hand side is the accumulated shot-noise multiplied by the SNR, and $e$ is the fundamental unit of electric charge. Solving this yields a measurement integration time of $\tau_m \sim 10^{-3}~\textrm{s}$.

In order to complete the measurement analysis we need to address the second issue identified above and confirm that the state of the nuclear spin does not flip within the measurement time -- i.e. the measurement time $\tau_m$ has to be shorter then the nuclear spin flip time $T_1$. Once the 2DEG current is switched on (the 2DEG current is off during the electron-nucleus state transfer) the dynamics of the donor electron due to scattering and microwave driving will contribute to the decoherence of the nuclear spin. Donor nuclear spin relaxation is not well characterized under these conditions but we expect that in large magnetic fields the donor electron dynamics contributes primarily only to dephasing of the nuclear state. This can be made precise by performing perturbation theory on \erf{eq:ham} in the parameter $A/\Delta$, where $\Delta \equiv \omega_e - \omega_n = B(g_e\mu_B - g_n \mu_n)$. In the \textit{detuned} regime where $A/\Delta \ll 1$, the effective Hamiltonian describing the coupled systems is 
\beq
H \approx H_{\textrm{eff}} = \frac{1}{2}\omega_e \sigma_z^e - \frac{1}{2}\omega_n\sigma_z^n + A\sigma_z^e\sigma_z^n + \frac{A^2}{\Delta}(\sigma_z^e - \sigma_z^n)
\eeq
This effective Hamiltonian is of course also the justification for the secular approximation made earlier (see below \erf{eq:ham}). Therefore we see that to first order in $A/\Delta$ the donor electron can only dephase the nuclear spin and that direct contributions to nuclear spin $T_1$ through the hyperfine interaction are small. Secondary mechanisms such as phonon-assisted cross relaxation can, in principle, also contribute to the nuclear $T_1$ during electron driving. However, these contributions were shown to be very small by Feher and Gere who demonstrated that the electron-nuclear cross relaxation time, $T_x$, under electron driving conditions is on the order of hours \cite{Feh.Ger-1959}. Given this extremely long cross-relaxation time and the equivalently long nuclear $T_1$ in a static electron environment at low temperatures \cite{Feh.Ger-1959, Tyr.Mor.etal-2006}, we conclude that the nuclear $T_1$ in the presence of electron driving will be comfortably larger than $\tau_m \sim 10^{-3}~\textrm{s}$. Indeed, this has very recently been confirmed by explicit measurements of nuclear $T_1$ under the conditions of electron driving \cite{Lyo-2008}.  This analysis implies that once the measurement collapses onto a nuclear basis state, the nuclear spin state does indeed effectively remain there and therefore the EDMR measurement satisfies the QND requirement on the qubit state.

\section{Conclusion}
\label{sec:conc}

By utilizing resonant pulse gates and 2DEG-mediated-EDMR readout, we have proposed a realistic scheme for measuring the spin state of a single donor electron in silicon. By making use of the hyperfine coupled donor nuclear spin, the readout scheme provides a single shot measurement that  is both projective and QND.  The QND aspect also makes this technique an effective method for initializing the state of the nuclear spin. 

We have analyzed the measurement procedure, the factors which influence the signal-to-noise ratio, and the experimental apparatus to arrive at realistic modifications/improvements that can be made to current 2DEG-based EDMR apparati \cite{Lo.Bok.etal-2007} so that a single nuclear spin can be measured. One concern is that at the required transistor size of $\sim 30\times 30 \textrm{nm}^2$, the MOSFET device will no longer act as a 2DEG but rather more like a quantum dot.  However, while the proposed MOSFET is small, there are no physical tunnel barriers between the transistor island and the source and drain leads (no intentional confinement).  Important Coulomb blockade effects have indeed been observed in such MOSFETs, but only in a biasing regime with very low source-drain biases \cite{San.Spe.etal-1999, Par.Lid.etal-2004, Fuj.Ino.etal-2006} and in sub-threshold currents at very low gate biases \cite{Sel.Lan.etal-2006}.  In these studies, confinement was found to be induced by impurity potentials \cite{San.Spe.etal-1999, Fuj.Ino.etal-2006}.   It has also been shown that one can easily tune in and out of the Coulomb blockade regime by suitable gate and source-drain biasing \cite{San.Spe.etal-1999} (and see also Ref. \cite{Fuj.Ino.etal-2006} for related results with tunable tunnel barriers). In the experiments we envision, which are closely related to and inspired by recent demonstrations of spin dependent transport in micron-scale devices \cite{Gho.Sil-1992, Lo.Bok.etal-2007}, the source-drain as well as the gate bias can be tuned over a large range of voltages, from the very low values needed to study interesting and important Coulomb blockade effects in a low current regime ($< 50$ nA) to higher bias values desirable for electrical detection of magnetic resonance through scattering of conduction electrons off neutral donors \cite{Gho.Sil-1992, Lo.Bok.etal-2007} with much higher channel currents ($\sim 1 ~\mu$A). Such MOSFET devices can be operated as 2DEGs and away from the quantum dot regime, by imposing large enough source-drain and gate biases.

Finally, we note that the fact that the measurement is facilitated by the nucleus of the donor atom intimates a hybrid donor qubit where quantum operations are carried out on the electron spin and the state is transfered to the nucleus for measurement and storage (advantageous due to the longer relaxation times). Although the above analysis was done with the example of a phosphorous donor, it applies equally well to other donors, such as antimony \cite{Sch.Lid.etal-2006, Lo.Bok.etal-2007}, and some paramagnetic centers \cite{Fuc.Dob.etal-2008}.  One merely has to isolate two (dipole-transition allowed) nuclear spin levels to serve as qubit basis states and transfer the electron state to these nuclear states with resonant pulses as outlined here. 

\section{Appendix}
Here we detail the procedure used in the main text for calculating the amount of time taken for $n$ scattering events. We assume a MOSFET channel of width $W=30$nm probed with a current of $1\mu$A. This means that there are $I/e \approx 1.6\times 10^{13}$ electrons per second crossing the channel. However, to gain a more accurate estimate of the number of electrons interrogating the donor electron per second we scale this total number by a ratio of the scattering length to the width of the channel: $\sqrt{\sigma}/W$. A crude estimate of the scattering cross section, $\sigma$, can be obtained using the singlet and triplet scattering lengths given in Ref. \cite{Hon-1966}: $a_s = 6.167 \AA$ and $a_t = 2.33 \AA$. These scattering lengths are for three dimensional electron-hydrogen scattering in bulk-doped systems, however we consider them sufficient for an order of magnitude estimate of the number of interrogating electrons per second. In terms of these scattering lengths, the scattering cross section for the 2DEG interacting with an ensemble of donors is \cite{Gho.Sil-1992}:
\bqa
\sigma = 2\pi \left[ (a_s^2 + 3a_t^2) - (a_s^2-a_t^2)P_c^0 P_i^0 \right]
\label{eq:scattering_cs}
\eqa
where $P_c^0$ and $P_i^0$ are the conduction band and impurity polarization. Since $P_c^0 \ll 1$ we will approximate the above expression as:
\bqa
\sigma \approx 2\pi (a_s^2 + 3a_t^2) = 341.3 ~\AA^2
\label{eq:approx_cs}
\eqa
For a single donor, \erf{eq:scattering_cs} should be modified to take into account the time-dependent ``polarization" of the single donor spin (see Sec. \ref{sec:sens}). However, since this quantity drops out in the final approximate expression  for the cross section, \erf{eq:approx_cs}, we will not be concerned with this modification. 

Using this estimate of the average scattering cross section, the number of electrons interrogating the donor per second is given by
\beq
n_e \approx 1.6 \times 10^{13} \cdot ~\frac{18.5 \times 10^{-10}}{30 \times 10^{-9}} \approx 1 \times 10^{12}
\eeq

Therefore we estimate that the time taken for $n$ scattering events is $n\times10^{-12}$ seconds.

\begin{acknowledgments}
We thank NSA (Grant MOD713106A) for financial support. MS and KBW were also supported by NSF (Grant EIA-0205641) and TS by DOE (Contract DE-AC02-05CH11231).  We are grateful to S. Lyon, A. M. Tyryshkin, C. C. Lo and J. Bokor for helpful discussions.
\end{acknowledgments}

\bibliography{/Users/mohan/Documents/bibdesk/mybib}

\end{document}